\documentclass[]{appolb}
\usepackage{graphicx}

\begin{document}

\def\Journal#1#2#3#4{{#1} {\bf #2}, #3 (#4)}

\def\NCA{Nuovo Cimento}
\def\NIM{Nucl. Instr. Meth.}
\def\NIMA{{Nucl. Instr. Meth.} A}
\def\NPB{{Nucl. Phys.} B}
\def\NPA{{Nucl. Phys.} A}
\def\PLB{{Phys. Lett.}  B}
\def\PRL{{Phys. Rev. Lett.}}
\def\PRC{{Phys. Rev.} C}
\def\PRD{{Phys. Rev.} D}
\def\ZPC{{Z. Phys.} C}
\def\JPG{{J. Phys.} G}
\def\EPJ{{Eur. Phys. J.} C}
\def\EPJST{{Eur. Phys. J.} - Special Topics}
\def\RPP{{Rep. Prog. Phys.}}

\title{Heavy Flavor Results at RHIC \\ 
- A Comparative Overview%
\thanks{Presented at Strangeness at Quark Matter 2011 conference. Send any remarks to {\tt XDong@lbl.gov}}%
}
\author{Xin Dong
\address{Nuclear Science Division, Lawrence Berkeley National Lab, \\
  MS70R0319, One Cyclotron Road, Berkeley, CA 94720, USA}
}
\maketitle
\begin{abstract}
I review the latest heavy flavor measurements at RHIC
experiments. Measurements from RHIC  together with
preliminary results from LHC offer us an opportunity to
systematically study the sQGP medium properties. In the end, I will outlook a
prospective future on precision heavy flavor measurements with detector
upgrades at RHIC.
\end{abstract}
  
\section{Introduction}
Heavy quarks are expected to be a clean
and penetrating probe to study the sQGP matter created in
heavy ion collisions because of its intrinsic large mass property.
By studying interactions between heavy quarks and medium, one can learn in
detail the flavor dependence of parton energy loss mechanism, and also
the medium's degree of thermalization by looking at the medium response
to heavy quarks. Precision measurements on heavy flavor hadrons in a
wide kinematic region will be unique to understand these details. Full
reconstruction of charm hadrons has significant advantages over
semi-leptonic decay leptons because of complete kinematics and clean
interpretations.

Quarkonium suppression due to color screening has been originally
proposed as a smoking gun for QGP formation. But various cold and hot
medium effects complex this story and there hasn't been a direct
evidence of color screening so far. Bottomonium (or
$\Upsilon$) production at RHIC may offer us a unique opportunity to
directly observe this signature because many cold and hot medium
effects (cold nuclear absorption, regeneration etc.) are expected to
negligible at RHIC. While at LHC, these become complicated, particularly the
regeneration process for bottomonium production can be significant,
which introduces difficulties in the interpretation.


\section{Latest RHIC Results}
In this proceedings, I would like to focus on latest RHIC results on charm
hadron and quarkonium ($J/\psi$, $\Upsilon$) measurements. 

\subsection{Charm hadron measurements}
\begin{figure}\centering{
\resizebox{0.545\columnwidth}{!}{
  \includegraphics{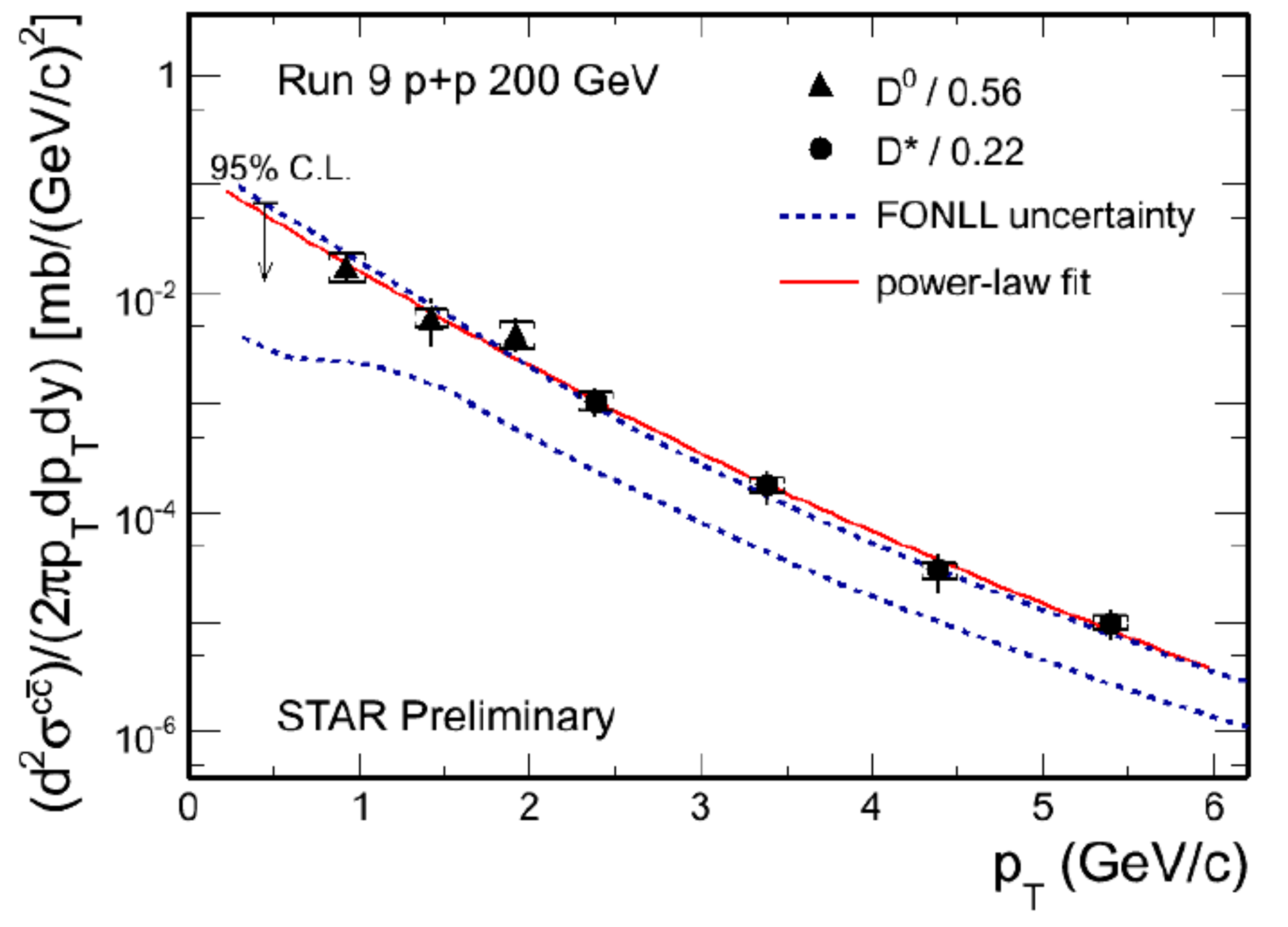}}
\resizebox{0.415\columnwidth}{!}{
  \includegraphics{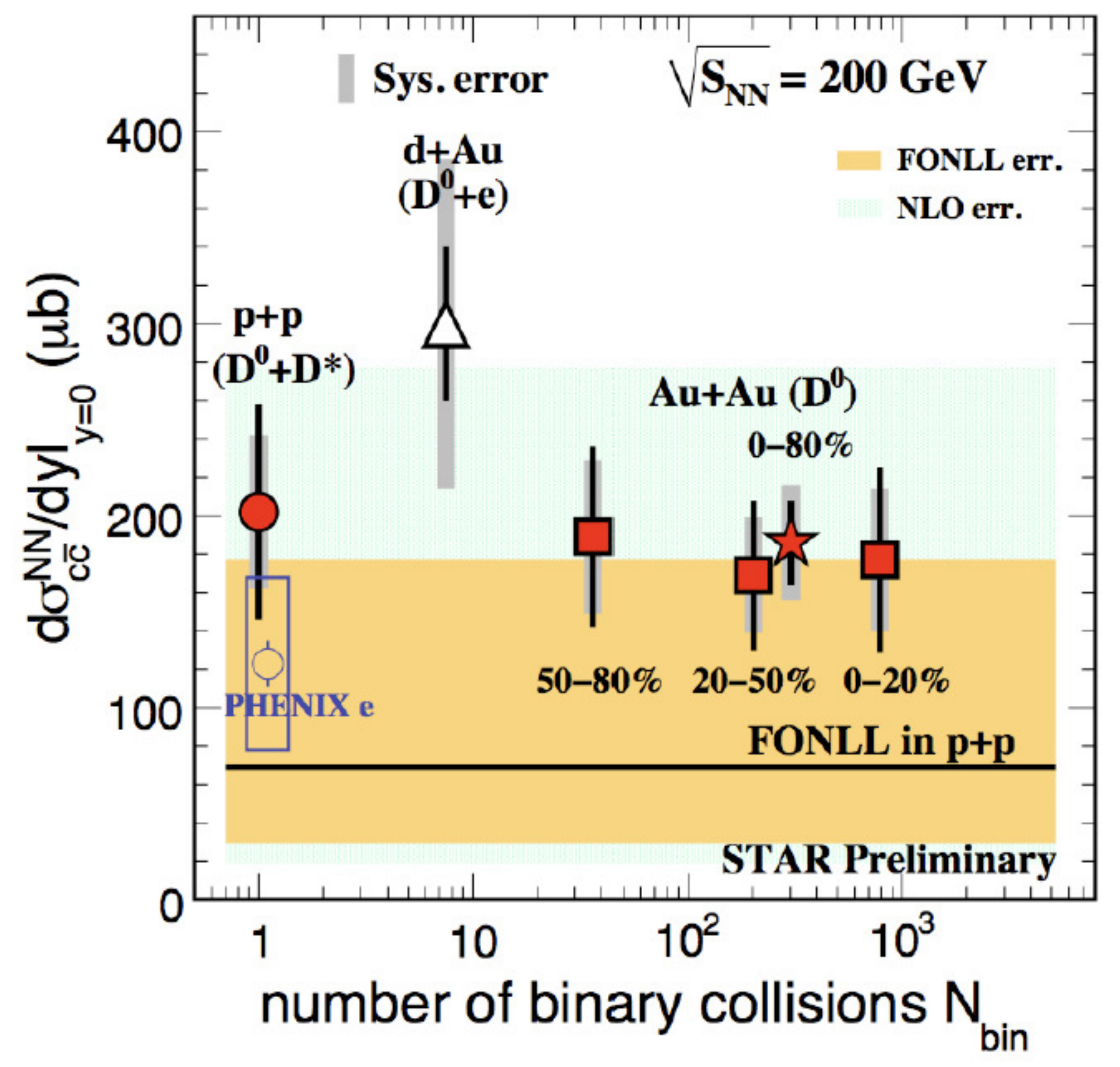}}
}
\caption{Left: Charm hadron ($D^0$ and $D^{*+}$) production cross
  section in $p+p$ collisions at $\sqrt{s}$ = 200 GeV from
  STAR. Measurements are consistent with the upper bound of the FONLL
  pQCD calculation. Right:Total charm production at mid-rapidity per
  nucleon-nucleon collisions from $p+p$ to central Au+Au collisions
  from STAR. The measurements demonstrate approximate number of binary
scaling for the total charm production cross section.}
\label{fig:1}       
\end{figure}

Figure~\ref{fig:1} shows the recent STAR measurements on the charm
hadron production cross sections in $p+p$ and minimum bias Au+Au
collisions~\cite{ZhangQM}. Charm hadrons were reconstructed
via hadronic decays. Although it was not possible to reconstruct
secondary decay vertices for charm hadrons in these measurements, STAR managed
to overcome the large combinatorial background with large amount of
statistics. The left plot shows the mid-rapidity production cross
sections for $D^0$ and $D^{*+}$ scaled to $c\bar{c}$ pairs vs. $p_T$
and the result is compared to a pQCD FONLL
calculation~\cite{FONLL}. One can see over a wide $p_T$ region, the
measurement is consistent with the upper bound of this FONLL
calculation. An interesting finding is that the charm hadron cross
sections measured by CDF~\cite{CDFcharm} and ALICE~\cite{ALICEcharm}
at higher energies up to 7 TeV are also closer to the upper limits of FONLL
calculations. 
Similarly, the cross section of non-photonic
electrons is also consistent with the upper bound of FONLL
calculations at $p_T(e)>$ 1 GeV/$c$~\cite{STARnpepp,PHENIXnpepp}. 
The STAR $D$ meson measurement covers about 70\% of total
$p_T$ acceptance, leading to a reasonable constrain to the total charm
cross section. Fig.~\ref{fig:1} right plot shows the total charm cross
section per nucleon-nucleon collisions at mid-rapidity from $p+p$ to
central Au+Au collisions. Latest STAR Au+Au results were extracted
from the $D^0$ spectrum measurements covering $p_T$ from 0.4 up to
$\sim$ 5 GeV/$c$. The $c\bar{c}$ cross sections were obtained
assuming the $c\rightarrow D^0$ fragmentation ratio still holds in
Au+Au collisions which need to be tested in future measurements. The
results exhibit an approximate $N_{bin}$ scaling indicating charm
quarks are predominantly produced from initial hard scatterings.

\begin{figure}\centering{
\resizebox{0.47\columnwidth}{!}{
  \includegraphics{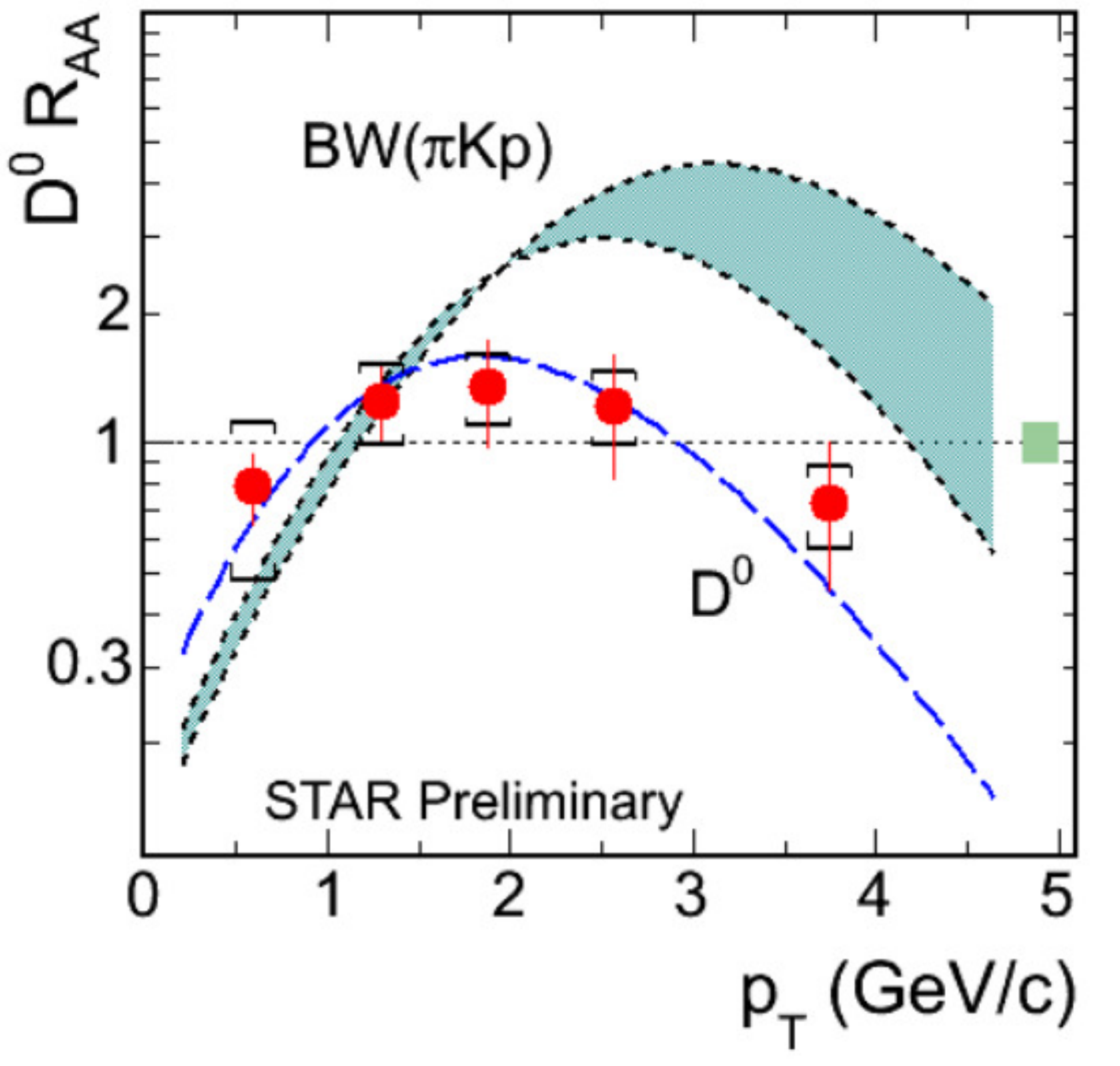}}
\resizebox{0.48\columnwidth}{!}{
  \includegraphics{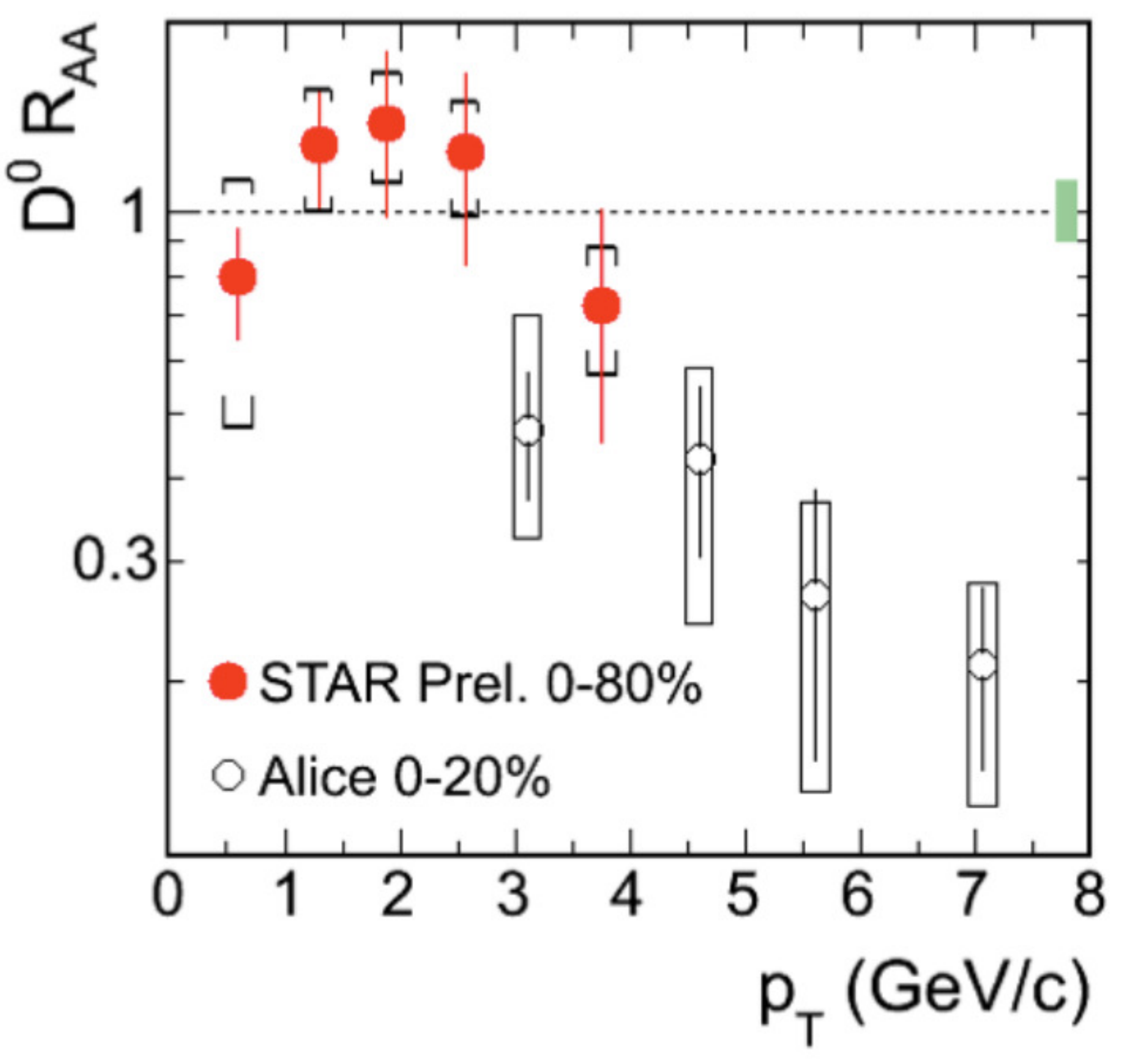}}
}
\caption{Left: $D^0$ meson $R_{AA}$ in 0-80\% minimum bias Au+Au
  collisions at RHIC from STAR. The blue dashed line depicts the
  $R_{AA}$ with the Au+Au $D^0$ spectrum fit to the Blast-Wave model. The shaded
  area depicts the $R_{AA}$ with Au+Au spectrum calculated in the
  Blast-Wave model with light hadron freeze-out parameters. Right: $D^0$ meson
  $R_{AA}$ in 0-80\% Au+Au collisions at $\sqrt{s_{NN}}$ = 200 GeV from STAR and $D$ meson
  $R_{AA}$ in 0-20\% Pb+Pb collisions at $\sqrt{s_{NN}}$ = 2.76 TeV
  from ALICE.}
\label{fig:2}       
\end{figure}

The $D^0$ spectrum in Au+Au collisions can be compared to the
reference $p+p$ data to calculate the nuclear modification factor
$R_{AA}$. Figure~\ref{fig:2} left panel shows the recent results from the STAR
measurements. The $R_{AA}$ covers $p_T$ up to $\sim$ 5
GeV/$c$. Although the values are consistent with unity given current
uncertainties, there seems to be a hint of modification for $D^0$
production in this $p_T$ region. The blue dashed line depicts the
$R_{AA}$ calculated based on the Blast-Wave fit to the Au+Au
spectrum. If one uses the Blast-Wave model with light hadron
freeze-out parameters and calculates the $R_{AA}$, the result is shown
as shaded area in the plot. The data points are significantly
different from this predication indicating the charm hadrons
freeze-out differently from the system compared to light hadrons. When
going towards higher $p_T$, the data points tends to show a slight
suppression, however, the uncertainties are large. The right panel
of Fig.~\ref{fig:2} plots the $D^0$ $R_{AA}$ from STAR together with
the high $p_T$ measurement from central
(0-20\%) Pb+Pb collisions by ALICE~\cite{ALICEcharm}. One should note
the difference in the centrality for these two sets of data
points. After considering this, there seems to be a consistent trend
of $D^0$ $R_{AA}$ between RHIC and LHC although the uncertainties need
to be shrink for detailed investigation. One should be
aware that $D^0$ $R_{AA}$ less than unity doesn't necessarily mean the
suppression of charm quark production in heavy ion collisions. The
sQGP medium in heavy ion collision may modify the distributions of
charm quarks into different charm hadrons compared to $p+p$ collisions
because hadronization scheme other than fragmentation
(e.g. coalescence) can be significant as already been observed for
light flavor hadrons. A complete understanding of charm quark energy
loss requires measurements of all ground state charm hadrons
in heavy ion collisions.

\subsection{$J/\psi$ measurements}
$J/\psi$ production has been reported by the PHENIX collaboration,
focusing on the low $p_T$ suppression
observation~\cite{PHENIXjpsi}. With a couple of subsystem upgrades,
STAR has been able to measure $J/\psi$ production with improved
statistics, particularly with significant capability covering high
$p_T$ up to $\sim$ 10 GeV/$c$. A combination of both PHENIX and STAR
measurements covers significant large acceptance in
4$\pi$~\cite{STARjpsippCuCu,PHENIXjpsipp,TangQM}, offering us a great
opportunity to constrain the quarkonium production mechanism as well as
to learn the cold and hot medium properties. Most newly
developed models can produce the cross section data points within
accessible kinematic region. Precision cross section measurements
provide constrains on model calculations. In the meantime, more
differential measurements (e.g. polarization) and/or with more extended
kinematic coverage can allow us to disentangle different models and
pin down the quarkonium production mechanism.

\begin{figure}\centering{
\resizebox{0.48\columnwidth}{!}{
  \includegraphics{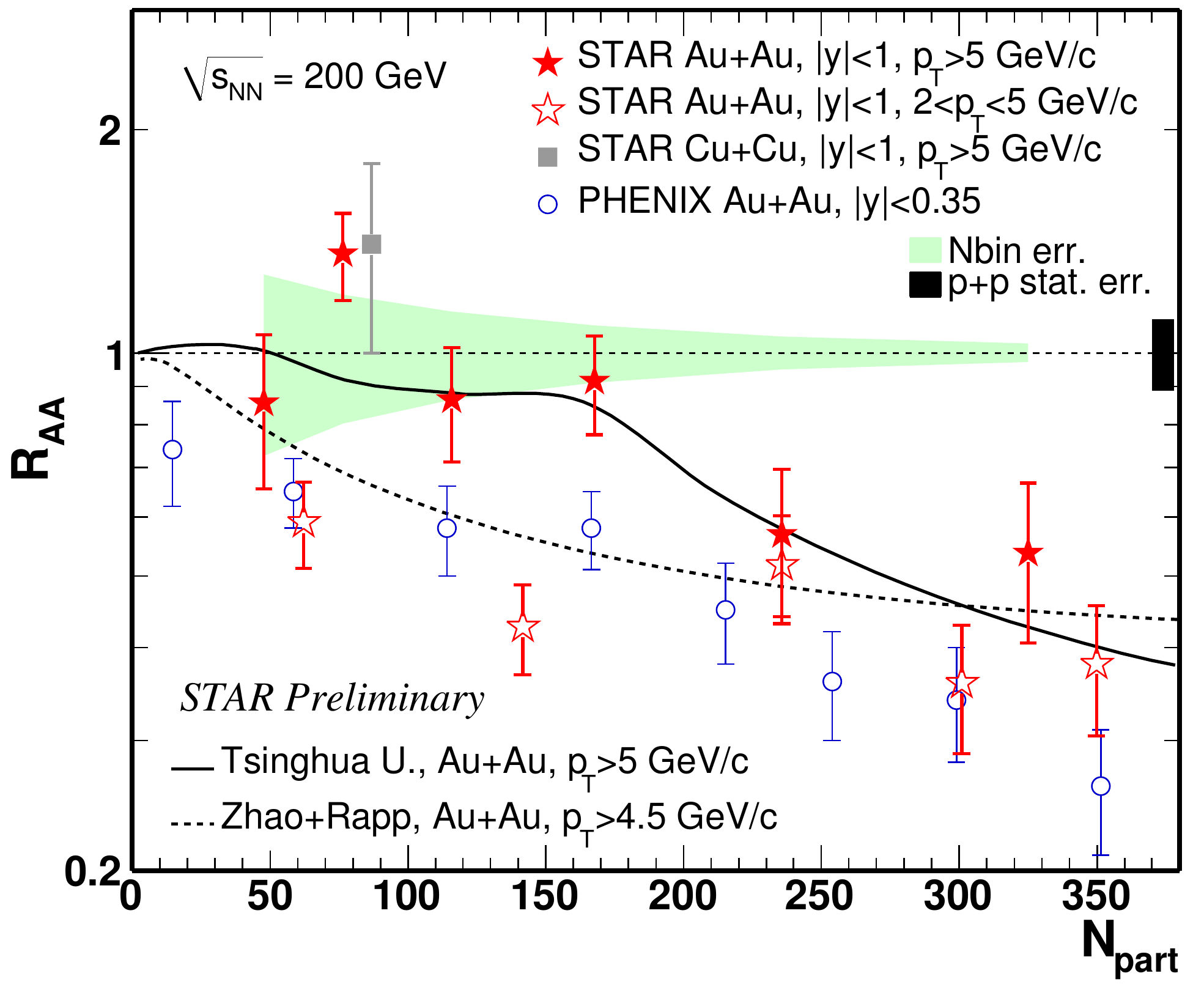}}
\resizebox{0.48\columnwidth}{0.40\columnwidth}{
  \includegraphics{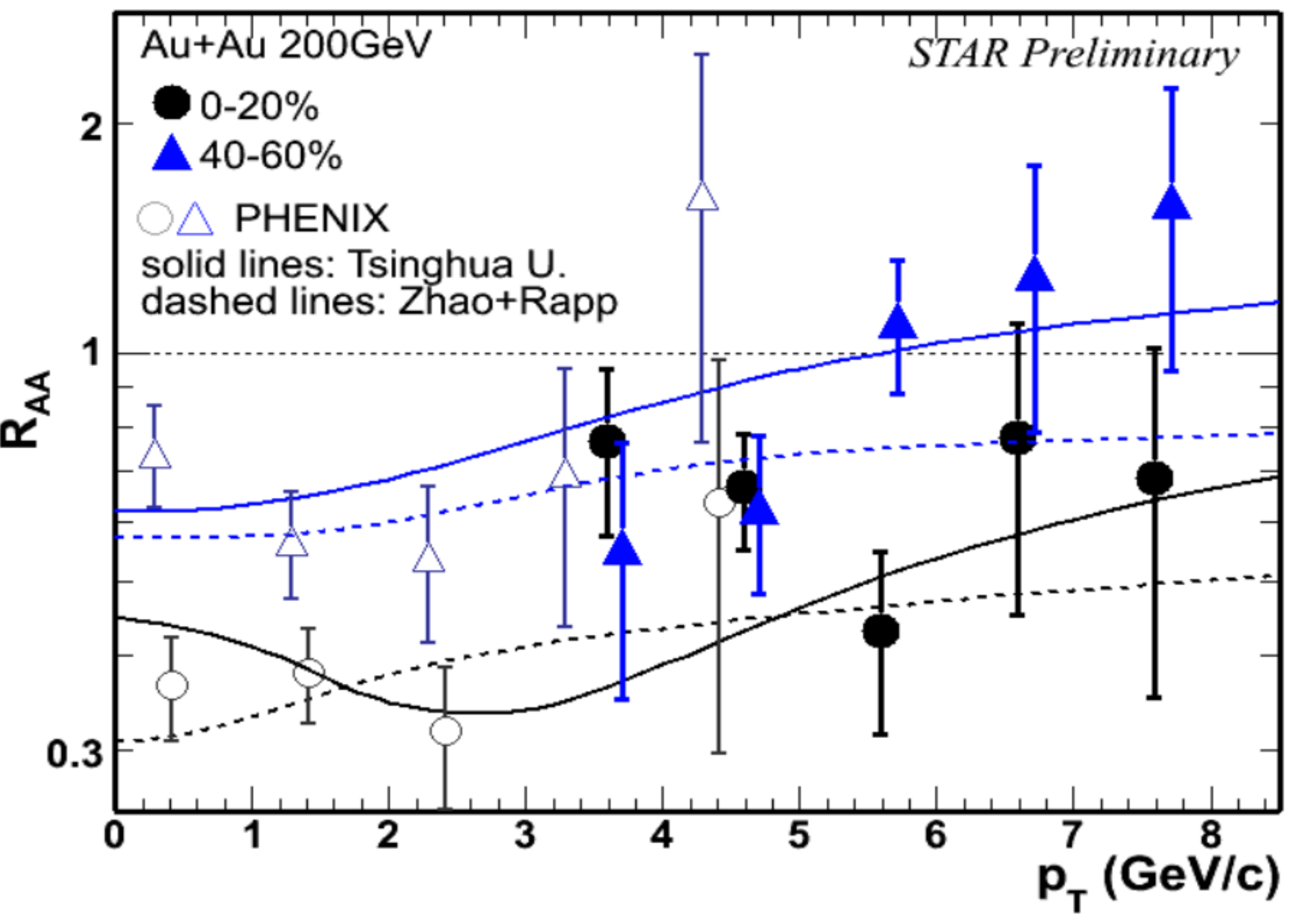}}
}
\caption{Left: Nuclear modification factor $R_{AA}$ of $J/\psi$
  vs. centrality. Measurements include PHENIX and STAR low $p_T$
  results in Au+Au collisions, STAR high $p_T$ results in Cu+Cu and
  Au+Au collisions. Curves shown in the plot are two model
  calculations which include both $J/\psi$ dissociation and
  recombination processes. Right: $R_{AA}$ of $J/\psi$ vs. $p_T$ in
  Au+Au collisions from RHIC measurements compared to model calculations.}
\label{fig:5}       
\end{figure}

With significantly improved statistics at high $p_T$, RHIC experiments
are now able to study the $p_T$ dependent $J/\psi$
$R_{AA}$~\cite{PHENIXjpsi,TangQM,PowellEPS}. Figure~\ref{fig:5} shows
the $J/\psi$ $R_{AA}$ results vs. centrality (left) and $p_T$ (right). On the left plot, the
latest STAR Au+Au results were divided into two $p_T$ bins. The low
$p_T$ (2-5 GeV/$c$) data show consistency with the PHENIX published
data points. High $p_T$ ($>$ 5 GeV/$c$) data show systematically
higher $R_{AA}$ compared to low $p_T$. In peripheral Au+Au
collisions, the high $p_T$ data are consistent with no
suppression, while in central collisions, $J/\psi$'s are still
significantly suppressed, which may be due to the color-screening
effect. On the right, the $R_{AA}$ values are plotted vs. $p_T$
for two centrality bins and one can see the $p_T$ dependent structure
more clearly. Also plotted on these two plots are theoretical model
calculations which both include the $J/\psi$
dissociation in the QGP phase as well as the regeneration
process~\cite{LiuJpsi,ZhaoJpsi}. Both model calculations generally describe the data
well. The system size and $p_T$ dependence of $J/\psi$ $R_{AA}$ can be
attributed to the formation time and/or leakage effects.

There have been many interesting results from LHC experiments
recently. ATLAS and CMS also observed large suppression of high $p_T$
$J/\psi$ in central Pb+Pb collisions~\cite{ATLASjpsi,CMSjpsi}, and the
suppression seems to be stronger than that of RHIC high $p_T$
$J/\psi$'s, consistent with more suppression in a larger size
system. ALICE reported low $p_T$ $J/\psi$ $R_{AA}$ at forward
rapidity~\cite{ALICEjpsi}. Comparing to the PHENIX result, the higher
$R_{AA}$ values observed is consistent with more regeneration
at LHC than at RHIC. Although current observations seem to
qualitatively agree with expectations, but quantitatively, we need
further systematic studies on both RHIC and LHC to understand
the quarkonium production mechanism as well as the medium
properties. This also requires precision measurements in $p(d)$+A
collisions to control the shadowing and the cold nuclear absorption effects.

\begin{figure}\centering{
\resizebox{0.84\columnwidth}{!}{
  \includegraphics{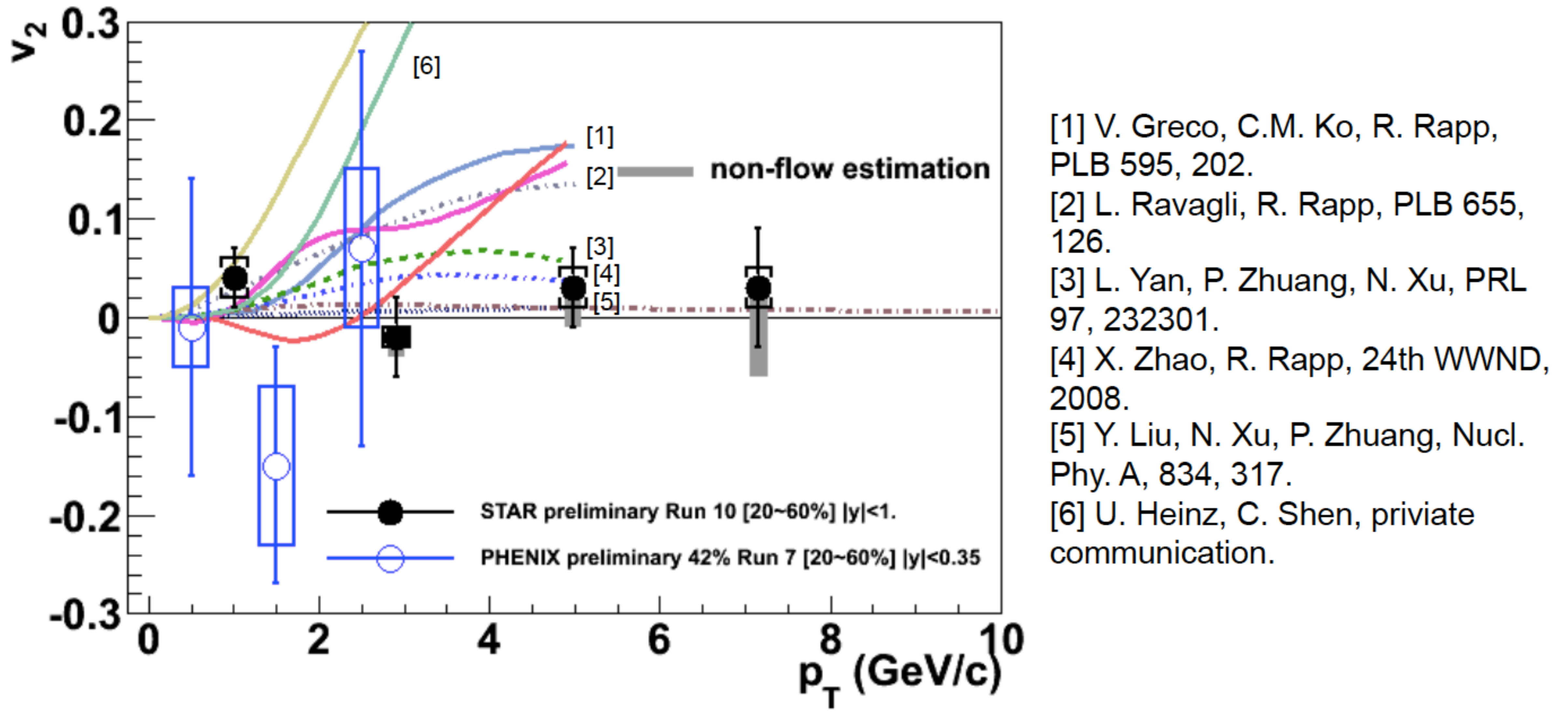}}
}
\caption{$J/\psi$ elliptic flow ($v_2$) measurements from RHIC
  experiments in 20-60\% Au+Au collisions compared with various model
  predictions.}
\label{fig:6}       
\end{figure}

Another striking result reported by STAR is the $J/\psi$ elliptic flow
($v_2$) measurement~\cite{QiuThis}. It has a significant improvement in term of
precision compared the previous PHENIX
measurement~\cite{PHENIXjpsiv2}. Figure~\ref{fig:6} shows the $J/\psi$
$v_2$ results from both PHENIX and STAR in 20-60\% Au+Au collisions
and they are compared to various model predictions. Data show at
$p_T>$ 2 GeV/$c$, there is no sizable $v_2$ for $J/\psi$, which
disfavors coalescence production from thermalized charm quarks. 

One should note the valid centrality and $p_T$ regions for this
statement. Based on model calculations~\cite{LiuJpsi} which generally
reproduce the $J/\psi$ $R_{AA}$ and $v_2$, the coalescence
contribution in 20-60\% Au+Au collision at RHIC is not dominant,
and it mostly contributes in low $p_T$ ($<$ 3 GeV/$c$) region. High
$p_T$ region is still dominated by the initial $J/\psi$ production
plus possible dissociation in the medium. To get insight of the clean
charm quark $v_2$ from the measurement of $J/\psi$ at RHIC energy, one
need to focus on the $p_T$ region below $\sim$ 3 GeV/$c$ and central
collisions. This requires further improvement in the experimental
precision. Another alternate way to learn the charm quark $v_2$ will
be to measure the charm hadron $v_2$ as experimentally, and it will be
achievable with great precision with help of the future silicon vertex detector.

\subsection{$\Upsilon$ measurements}
Study of bottomonium production at RHIC is a unique way to get insight
of the originally proposed QGP color screen signature. The challenge for
RHIC experiments is the statistics which makes this measurement likely to
be a multi-year program. Both PHENIX and STAR have reported
$\Upsilon$ (1S+2S+3S states if not specified) signals from $p+p$ and
Au+Au collisions since QM2009. The STAR
published and PHENIX preliminary $\Upsilon$ cross sections in $p+p$
collisions are consistent with pQCD model calculations and lie on the
trend of energy dependence curve~\cite{STARupsilonpp,PHENIXupsilon}. Recently, with significantly
improved statistics in Au+Au collisions, STAR was able to measure the
centrality dependence of $\Upsilon$ $R_{AA}$~\cite{ReedQM}.

\begin{figure}\centering{
\resizebox{0.56\columnwidth}{!}{
  \includegraphics{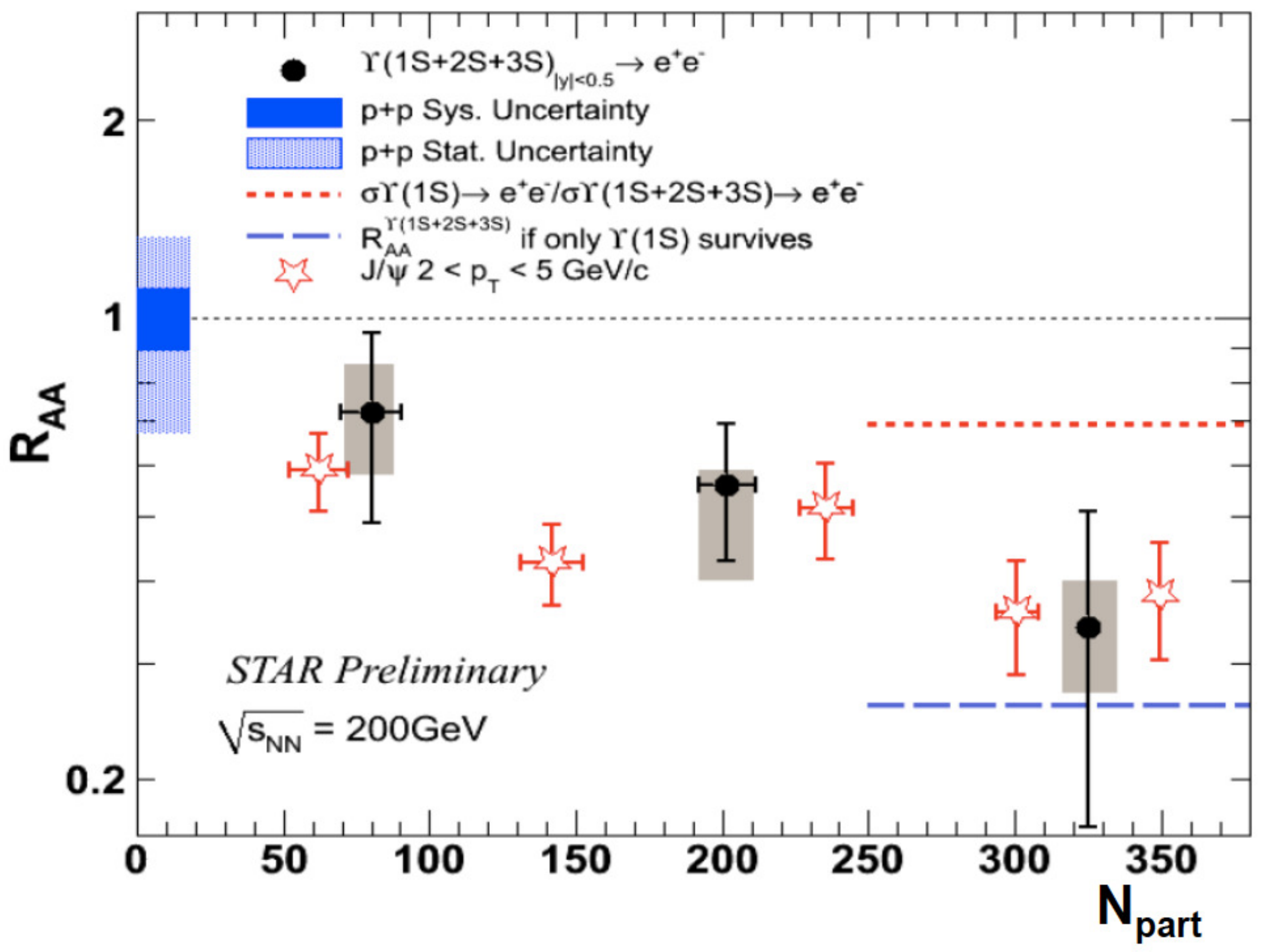}}
}
\caption{$\Upsilon$(1S+2S+3S) $R_{AA}$ vs. centrality from STAR
  compared to expectations from two naive $\Upsilon$ melting scenarios.}
\label{fig:7}       
\end{figure}

Figure~\ref{fig:7} shows the $\Upsilon$ $R_{AA}$
vs. centrality. It shows a trend of $\Upsilon$ suppression in central
Au+Au collisions. Taking into account all current uncertainties from
Au+Au and $p+p$ reference, the measured value is $\sim$ 3$\sigma$
below unity. The red dotted line depicts a naive expectation assuming
2S and 3S states completely melt based on pQCD cross sections and
PDG branching ratios. The blue dashed line depicts the expect value
when all excited bottomonium states (including $\chi_b$) all melt and
only $\Upsilon$(1S) survies~\cite{KolleggerThesis}. The data now seem
to favor the scenario that only $\Upsilon$(1S) survies, but to learn
quantitatively suppression levels of different bottomonium states
requires significantly improved statistics in the future.

With great mass resolution, CMS collaboration was managed to separate 
(2S+3S) states from 1S state and they reported the separate $R_{AA}$ 
values in minbias Pb+Pb collisions at 2.76 TeV~\cite{CMSupsilon}. When combining them 
together to compare with RHIC results, the expected $R_{AA}$ for 
central collisions should be less than 0.42, which is generally
in line with the STAR result. However, one 
should note that the cold and hot medium effects may be significantly 
different between RHIC and LHC.

\section{Summary and Outlook}
In a short summary, RHIC heavy flavor program has achieved many
significant results particularly in the past couple of years. It will
remain as one of the focuses of RHIC heavy ion program in the upcoming
RHIC II era. Both PHENIX and STAR are building significant detector
subsystem upgrades now or in the coming years, e.g. PHENIX VTX and
FVTX, STAR HFT and MTD upgrades etc. These are all aiming for precision
measurements of both open heavy flavor and quarkonium production at RHIC
II. With the LHC experiments ongoing, RHIC heavy flavor program will
be complementary and remain competitive in many aspects. Some of the
measurements will be unique at RHIC, including: a) high precision open
charm hadron measurements at low $p_T$ to address charm-medium
interactions, b) bottomonium production measurements as bottomonia are
expected to be clean at RHIC because of negligible contribution from
regeneration, c) heavy quark correlation measurements as heavy quarks are
expected to be back-to-back correlated in $p+p$ collisions which
allows clean interpretations for results in heavy ion collisions. These
systematic study of heavy flavor measurements at RHIC and LHC will
significantly improve our understanding of the sQGP matter by
quantifying its physical properties with controlled accuracy.


\end{document}